\documentclass[aps,prb,reprint,superscriptaddress,nofootinbib]{revtex4-2}

\usepackage{amsmath,amssymb,bm}
\usepackage{physics}
\usepackage{graphicx}
\usepackage{booktabs}
\usepackage{microtype}
\usepackage{hyperref}
\usepackage{xcolor}
\usepackage{siunitx}
\usepackage{multirow}

\hypersetup{hidelinks}

\newcommand{\D}{\mathcal{D}}
\newcommand{\Lio}{\mathcal{L}}
\newcommand{\ee}{\mathrm{e}}
\newcommand{\ii}{\mathrm{i}}
\newcommand{\kb}{k_{\mathrm B}}

\begin{document}

\title{Preparation-protocol-dependent quantum Mpemba dynamics in a magnetically tunable graphene nanotorus qubit}

\author{J. Furtado}
\email{job.furtado@ufca.edu.br}
\affiliation{Centro de Ci\^encias e Tecnologia, Universidade Federal do Cariri, 63048-080, Juazeiro do Norte, Cear\'a, Brazil}

\date{September 11, 2026}

\begin{abstract}
We investigate the quantum Mpemba effect in a graphene nanotorus qubit and show that its strength depends strongly on how the initial state is prepared. We compare two protocols that share the same final Hamiltonian, thermal bath, and Markovian Liouvillian: bare Gibbs preparation, in which the coherent drive is switched on only at the final quench, and driven steady-state preparation, in which the initial states are stationary states of the driven dynamics. Using the distance to the final stationary state and an integrated Mpemba parameter, we find broad regions of strong anomalous relaxation for bare Gibbs preparation, with $M_B$ approaching unity, whereas the driven steady-state protocol almost completely suppresses the effect. A Liouvillian-mode analysis reveals the mechanism: the bare Gibbs protocol typically yields a smaller hot-state weight in the slow active sector than the cold state, $R_s<1$, while the driven steady-state protocol predominantly gives $R_s>1$. Since the final Liouvillian spectrum is identical for both protocols at corresponding parameter points, the contrast originates from preparation-dependent modal weights rather than from changes in the decay-rate hierarchy. These results identify state preparation as a control parameter for anomalous quantum relaxation in a curved graphene qubit.
\end{abstract}

\maketitle

\section{Introduction}
\label{sec:intro}

The Mpemba effect describes the counterintuitive situation in which a system initially farther from equilibrium can approach a common stationary state faster than another system that begins closer to equilibrium. Its quantum counterpart can be formulated in terms of relaxation generated by an open-system Liouvillian, where the occurrence and strength of the effect depend not only on the decay spectrum but also on the projection of the initial state onto the corresponding decay modes~\cite{FedericoCarolo:21,Zhang2025,Furtado:2024pii}. In particular, a strong quantum Mpemba effect (QMpE) can occur when the initially hotter state has a vanishing or strongly suppressed component along the slowest-decaying Liouvillian mode.

The QMpE has been explored in a variety of Markovian many-body settings, including the Dicke model~\cite{FedericoCarolo:21}, the Anderson model describing quantum-dot reservoirs~\cite{Chatterjee:23}, the Lieb-Liniger model~\cite{Rylands:23}, and spin-chain systems~\cite{Kochsiek:22,Ares:23}. To the best of our knowledge, the first experimental observation of this phenomenon was reported only recently by J. Zhang \textit{et al}.~\cite{Zhang:24}, using a single trapped-ion platform in which a three-level system was encoded in the low-lying electronic states of a trapped $^{40}$Ca$^{+}$ ion. Despite this progress, these studies, together with several related works~\cite{Manikandan:21,Rylands:23,Ivander:23}, have generally focused on open-system environments whose effective reservoirs are generated by physical mechanisms different from those associated with genuine thermal baths at a well-defined finite temperature, \(T>0\). To address the corresponding issues of characterization and thermodynamic interpretation, alternative notions of temperature for isothermal processes have been introduced in the literature~\cite{Chatterjee:23,Chatterjee:23-b}, with particular emphasis on the analysis of heat exchange and entropy changes during the evolution.

More recent developments have broadened the scope of the QMpE beyond the identification of anomalous relaxation itself, placing increasing emphasis on its dynamical control and microscopic mechanisms. In Markovian settings, the role of the Liouvillian spectrum and of the initial-state projections onto slow decay modes has been further clarified in studies of intrinsic QMpE and quantum-circuit implementations~\cite{Qian:2024dde}, as well as in constrained Rydberg chains, where exact slow-mode selection was shown to provide a direct route to strong QMpE behavior~\cite{Xu:2026muy}. Related work has demonstrated that conserved quantities can enable the effect in weakly open quantum systems ~\cite{Ulcakar:2025mle}, while long-range interacting platforms have revealed a high degree of tunability of the relaxation inversion through external control parameters~\cite{Hallam:2025axu}. On the experimental side, the direct observation of the QMpE without bath engineering has further emphasized that anomalous relaxation can arise under comparatively natural dissipative conditions ~\cite{Chatterjee:2025vfn}. Complementary approaches have also focused on the detection of Mpemba behavior through experimentally accessible observables in open quantum systems~\cite{Bagui:2025fkn}. These advances have been accompanied by recent reviews that consolidate the rapidly expanding quantum Mpemba literature~\cite{Ares:2025onj} and place it within the broader context of accelerated nonequilibrium thermal relaxation and related Mpemba phenomena ~\cite{Teza:2025azr}.

A useful global witness was introduced in Ref.~\cite{Furtado:2024pii}. For two initial states, labelled hot and cold, evolving under the same final dynamical generator, one compares their distances $B_h(t)$ and $B_c(t)$ from the final stationary state. If $B_h(0)>B_c(0)$ but $B_h(t)<B_c(t)$ during part of the evolution, a quantum Mpemba crossing occurs. The associated parameter $M_B$ measures the integrated weight of the interval in which the initially hotter state is closer to the target than the initially colder state. This formulation is particularly convenient for physical qubit platforms because it can be applied directly to a device-specific Hamiltonian and dissipative model.

The graphene nanotorus qubit provides such a platform. In the previously proposed encoding, an electron confined to the inner region of a graphene nanotorus forms curvature-induced bound states. A static magnetic field modifies the confinement potential and allows the states $\ket{l=0,m=0}$ and $\ket{l=1,m=0}$ to define an effective two-level system, while a local oscillating electric field drives arbitrary single-qubit rotations~\cite{Furtado:2022uvk}. For a representative geometry with minor radius $r=350$~\AA, major radius $R=900$~\AA, and magnetic field $B=0.45$~T, the transition frequency is $\omega/2\pi\simeq25.9$~GHz~\cite{Furtado:2022uvk}. The same proposal explicitly identified environmental coupling, electron--phonon processes, hot-electron relaxation, and decoherence as important open problems for the platform.

Previous studies have established that strong QMpE behavior can be induced by engineering initial states or suppressing their overlap with slow Liouvillian modes. Here we address a distinct question: rather than optimizing arbitrary initial states, we compare two physically motivated preparation protocols that generate different density matrices before an otherwise identical final evolution. The present work focuses on a question that emerged naturally once the open-system dynamics was implemented: how strongly does the QMpE depend on the way the hot and cold states are prepared? This issue is distinct from a memory of microscopic history. Within the Markovian description used here, two identical density matrices evolved under the same Liouvillian necessarily follow the same trajectory. The preparation protocol matters because different procedures generate different initial density matrices, and therefore different overlaps with the decay modes of the final generator. This distinction turns out to be quantitatively decisive in the nanotorus model.

We compare two protocols. In the \emph{bare Gibbs protocol}, hot and cold Gibbs states are prepared with respect to the undriven qubit Hamiltonian and the coherent electric drive is activated at the final quench. In the \emph{driven steady-state protocol}, the drive is already present during preparation and the hot and cold states are the stationary states of their respective finite-temperature driven Liouvillians. Both protocols are then subjected to exactly the same final bath and final Hamiltonian. The comparison therefore isolates the effect of the initial-state structure.

The main numerical result is a striking separation between the two cases. Bare Gibbs preparation produces broad regions of strong QMpE, with $M_B$ reaching values extremely close to unity in several parameter scans. Driven steady-state preparation, by contrast, leaves $M_B$ close to zero over most of the same parameter space; when crossings occur, they are typically weak and delayed. The accompanying Liouvillian decomposition resolves the mechanism: although the final spectrum is the same for the two protocols at corresponding physical parameters, their initial projections onto the slow active sector are reversed, with the bare-Gibbs protocol favoring $R_s<1$ and the driven steady-state protocol predominantly yielding $R_s>1$. Thus the preparation sensitivity is not merely phenomenological but is directly encoded in how the initial states populate the decay modes of the common final generator.

The paper is organized as follows. Section~\ref{sec:torus} summarizes the effective nanotorus qubit. Section~\ref{sec:thermal} introduces the thermal Lindblad model and the two preparation protocols. Section~\ref{sec:baseline} establishes the no-drive baseline. Section~\ref{sec:witness} defines the Mpemba witness and the numerical procedure. Section~\ref{sec:results} presents the preparation-dependent numerical results. Section~\ref{sec:discussion} discusses the physical interpretation and limitations, and Sec.~\ref{sec:conclusion} summarizes the conclusions and outlook.

\section{Nanotorus qubit: minimal recap}
\label{sec:torus}

\subsection{Curvature-induced confinement and magnetic tuning}

\begin{figure*}[t]
\centering
\includegraphics[width=0.95\textwidth]{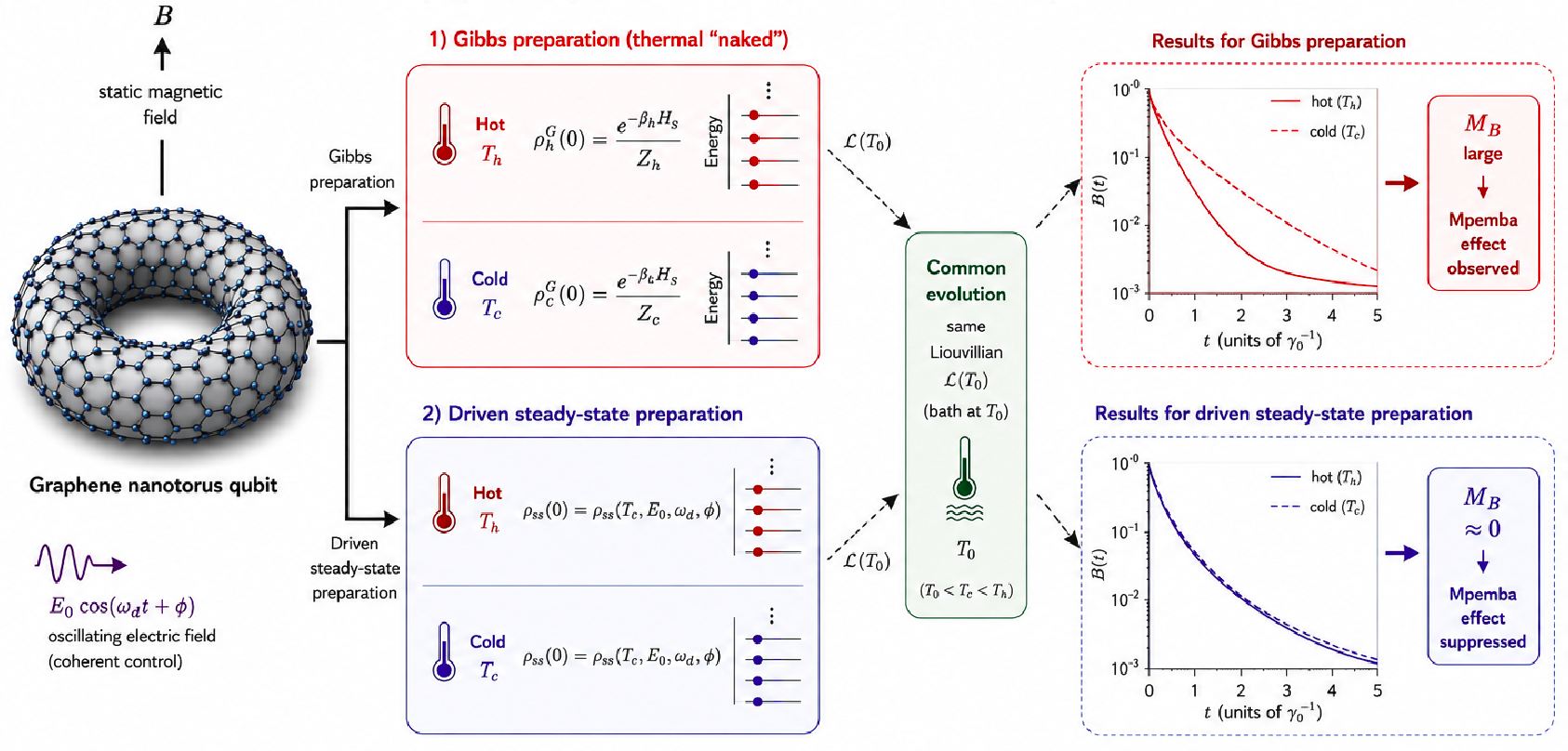}
\caption{Schematic of the nanotorus QMpE protocol. The static magnetic field $B$ fixes the bound-state structure and transition frequency, while the oscillating electric field provides coherent control. Hot and cold states are prepared at $T_h$ and $T_c$ and subsequently evolved under the same final Liouvillian associated with the colder bath $T_0$. The present work compares two ways of preparing $\rho_h(0)$ and $\rho_c(0)$: bare Gibbs states and driven stationary states.}
\label{fig:protocol}
\end{figure*}

We consider an electron confined to a graphene nanotorus of minor radius $r$ and major radius $R$. The axial symmetry allows the wave functions to be written as
\begin{equation}
 \psi_{n,l,m}(\theta,\phi)=\chi_{n,l}(\theta)\ee^{\ii m\phi},
\end{equation}
where $m$ is the angular-momentum quantum number along the symmetry axis and $\theta$ parametrizes the poloidal direction. For the nonrelativistic multilayer realization considered in Ref.~\cite{Furtado:2022uvk}, the effective one-dimensional Hamiltonian reads
\begin{equation}
 H(\theta)= -\frac{\hbar^2}{2m^*r^2}\frac{d^2}{d\theta^2}
 +V_{\rm bare}(\theta)+V_B(\theta)+V_E(\theta),
 \label{eq:fulltorusH}
\end{equation}
with $m^*=0.3m_0$ in the representative model. The bare contribution contains the da Costa curvature potential together with geometry-induced centrifugal terms. For a static magnetic field along the symmetry axis,
\begin{equation}
 V_B(\theta)=\frac{e^2B^2}{8m^*}(R+r\cos\theta)^2
 -\frac{me\hbar B}{2m^*}.
 \label{eq:VB}
\end{equation}
The magnetic field therefore reshapes the effective confinement potential rather than acting only as a conventional Zeeman shift.

For $m=0$, the field can be chosen such that two relevant low-lying bound states occupy the computational sector,
\begin{equation}
 \ket{0}\equiv\ket{l=0,m=0},\qquad
 \ket{1}\equiv\ket{l=1,m=0}.
 \label{eq:logical}
\end{equation}
After projection onto this subspace and subtraction of an irrelevant identity contribution, the bare qubit Hamiltonian is
\begin{equation}
 H_{\rm qb}=\hbar\omega(B,R,r)\,\sigma_+\sigma_-.
 \label{eq:Hqb}
\end{equation}
For the representative geometry adopted throughout this work, with
$r=350~\text{\AA}$, $R=900~\text{\AA}$, and $B=0.45~\text{T}$,
Ref.~\cite{Furtado:2022uvk} gives a qubit transition frequency of
$\omega/2\pi\simeq25.9~\text{GHz}$, corresponding to the characteristic
thermal scale $\hbar\omega/k_{\mathrm B}\simeq1.24~\text{K}$. This energy
scale provides a useful reference for choosing the hot, cold, and target
temperatures considered in the dissipative analysis below.

\subsection{Electric-field control}

The established control protocol uses a local oscillating electric field
\begin{equation}
 E(t)=E_0\cos(\omega_dt+\varphi),
 \label{eq:drive}
\end{equation}
which produces a projected coupling
\begin{equation}
 H_E=\hbar\Omega(E)(\sigma_++\sigma_-),\qquad
 \Omega(E)=\frac{\mu(B,R,r)E}{\hbar}.
 \label{eq:Omega}
\end{equation}
Near resonance and within the rotating-wave approximation, the driven qubit is described by
\begin{equation}
 H_{\rm eff}=\hbar\Delta\,\sigma_+\sigma_-
 +\frac{\hbar\Omega}{2}
 \left(\ee^{\ii\varphi}\sigma_-+\ee^{-\ii\varphi}\sigma_+\right),
 \label{eq:Heff_source}
\end{equation}
where $\Delta=\omega-\omega_d$ and $\Omega\equiv\Omega(E_0)$. This is the Hamiltonian used throughout the dissipative analysis.

\section{Thermal open-system model and preparation protocols}
\label{sec:thermal}

\subsection{Thermal Lindblad dynamics}

The original nanotorus proposal identified coupling to phonons, hot-electron relaxation, and environmental decoherence as unresolved ingredients of the qubit dynamics~\cite{Furtado:2022uvk,CastroNeto2009,Manes2007,Park2014,Hwang2013,Sedrakyan2021}. We therefore begin with the minimal Born--Markov master equation
\begin{align}
 \dot\rho={}&-\frac{\ii}{\hbar}[H_{\rm eff},\rho]
 +\Gamma_\downarrow(T)\D[\sigma_-]\rho
 +\Gamma_\uparrow(T)\D[\sigma_+]\rho
 \nonumber\\
 &+\Gamma_\phi(T)\D[\sigma_z]\rho,
 \label{eq:master}
\end{align}
where
\begin{equation}
 \D[L]\rho=L\rho L^\dagger-\frac{1}{2}\{L^\dagger L,\rho\}.
\end{equation}
For the numerical results below we use thermal bosonic rates
\begin{align}
 \Gamma_\downarrow(T)&=\gamma_0[n_B(\omega,T)+1],\\
 \Gamma_\uparrow(T)&=\gamma_0 n_B(\omega,T),
 \label{eq:thermalrates}
\end{align}
with
\begin{equation}
 n_B(\omega,T)=\frac{1}{\exp(\hbar\omega/\kb T)-1}.
 \label{eq:BE}
\end{equation}
The scale $\gamma_0$ is phenomenological at this stage. Unless otherwise stated, we use $\gamma_0/2\pi=1$~MHz and $\Gamma_\phi=0$. The absolute times therefore serve as a controlled dimensionless benchmark; a microscopic electron--phonon calculation is required before interpreting $\gamma_0$ as a device-specific relaxation rate.

\subsection{Bare Gibbs preparation}
\label{subsec:gibbs}

In the first protocol, the drive is absent during state preparation. The hot and cold states are Gibbs states of the bare qubit Hamiltonian,
\begin{equation}
 \rho_\nu^{\rm G}(0)=
 \frac{\exp[-H_{\rm qb}/(\kb T_\nu)]}
 {\Tr\exp[-H_{\rm qb}/(\kb T_\nu)]},
 \qquad \nu\in\{h,c\}.
 \label{eq:gibbsinitial}
\end{equation}
At $t=0$, the temperature is quenched to $T_0$ and the coherent electric drive is simultaneously activated. Both initial states then evolve under the same final Liouvillian $\Lio_0\equiv\Lio(T_0;E_0,B,R,r)$,
\begin{equation}
 \rho_{h,c}^{\rm G}(t)=\ee^{\Lio_0t}\rho_{h,c}^{\rm G}(0).
 \label{eq:gibbsevolution}
\end{equation}
This protocol therefore contains both a thermal quench and a coherent Hamiltonian quench.

\subsection{Driven steady-state preparation}
\label{subsec:steady}

In the second protocol the coherent drive is already present during preparation. The initial states satisfy
\begin{equation}
 \Lio(T_\nu;E_0,B,R,r)
 \left[\rho_\nu^{\rm ss}(0)\right]=0,
 \qquad \nu\in\{h,c\}.
 \label{eq:ssinitial}
\end{equation}
At $t=0$, only the bath temperature is changed to $T_0$, while the Hamiltonian and control parameters are kept fixed,
\begin{equation}
 \rho_{h,c}^{\rm ss}(t)=\ee^{\Lio_0t}\rho_{h,c}^{\rm ss}(0).
 \label{eq:ssevolution}
\end{equation}
Thus the two protocols have the same final generator but different initial states. This is the precise sense in which the QMpE is preparation-protocol dependent in the present model.

\section{No-drive baseline}
\label{sec:baseline}

Before discussing the numerical scans, consider $E_0=0$, so that $\Omega=0$, with $\Gamma_\phi=0$. For a thermal bath and an initial state diagonal in the energy basis, the excited-state population obeys
\begin{equation}
 \dot p=-\Gamma_1[p-p_0],\qquad
 \Gamma_1=\Gamma_\downarrow+\Gamma_\uparrow,
 \label{eq:populationeq}
\end{equation}
which gives
\begin{equation}
 p_\nu(t)-p_0=[p_\nu(0)-p_0]\ee^{-\Gamma_1t}.
 \label{eq:population}
\end{equation}
For the distance employed in Ref.~\cite{Furtado:2024pii}, the two trajectories inherit the same exponential factor,
\begin{equation}
 B_\nu(t)=B_\nu(0)\ee^{-\Gamma_1t}.
 \label{eq:distancebaseline}
\end{equation}
Hence
\begin{equation}
 B_h(0)>B_c(0)\;\Longrightarrow\;B_h(t)>B_c(t)
 \quad\forall t>0,
 \label{eq:nocrossing}
\end{equation}
and $M_B=0$. Numerically, the implementation reproduces this limit for both preparation protocols. Since the driven stationary state becomes the ordinary thermal state when $E_0=0$, the two preparation prescriptions also coincide in this limit. The no-drive result is therefore an important sanity check and confirms that the positive $M_B$ found below is genuinely associated with the driven multimode dynamics.

\section{Mpemba witness and numerical implementation}
\label{sec:witness}

\begin{figure*}[t]
\centering
\includegraphics[width=0.90\textwidth]{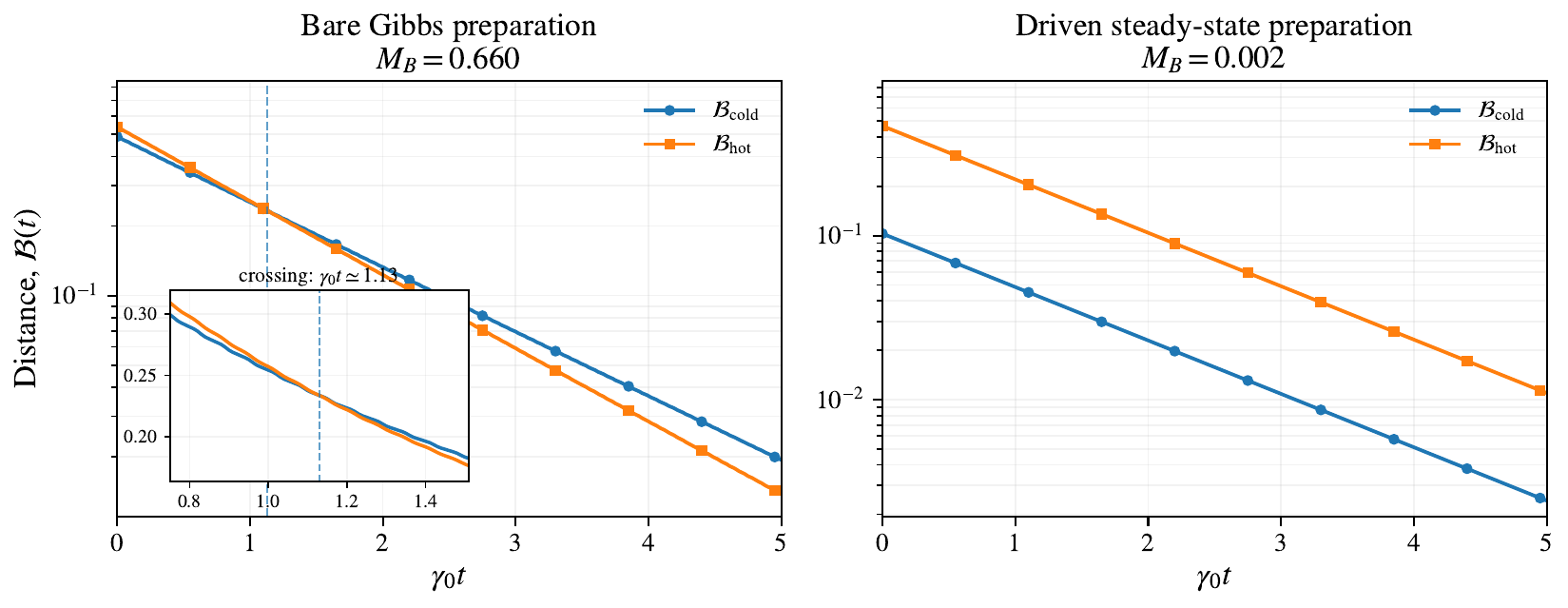}
\caption{Direct comparison of the distance dynamics at the reference parameter point. Left: bare Gibbs preparation gives $M_B=0.661$ and a clear crossing at $\gamma_0t\simeq1.13$. Right: driven steady-state preparation yields $M_B=0.002$; the crossing occurs only at $\gamma_0t\simeq11.1$, outside the plotted interval $0\le\gamma_0t\le5$. Both panels use the same final Liouvillian.}
\label{fig:protocoldecay}
\end{figure*}

Let $\rho_0$ denote the stationary state of the final Liouvillian at
temperature $T_0$. Following the distance convention employed in
Ref.~\cite{Furtado:2024pii}, we define
\begin{equation}
B_{h,c}(t)
=
\left[
\mathrm{Tr}
\left\{
\left[\rho_{h,c}(t)-\rho_0\right]^\dagger
\left[\rho_{h,c}(t)-\rho_0\right]
\right\}
\right]^{1/2},
\end{equation}
which corresponds to the Hilbert--Schmidt (Frobenius) distance between
the evolving state and the final stationary state. The QMpE occurs when the initial ordering
$B_h(0)>B_c(0)$ is reversed during the common relaxation dynamics, such
that $B_h(t)<B_c(t)$ over a finite time interval. The integrated Mpemba
parameter is then introduced as
\begin{equation}
 M_B=
 \frac{\displaystyle \int_{B_c>B_h}dt\,|B_c(t)-B_h(t)|}
 {\displaystyle \int_0^\tau dt\,|B_c(t)-B_h(t)|},
 \label{eq:MB}
\end{equation}
so that $0\le M_B<1$. The numerical integration uses $0\le\gamma_0t\le20$. The direct reference-point comparison in Fig.~\ref{fig:protocoldecay} is displayed over $0\le\gamma_0t\le5$. In the bare-Gibbs summary panel of Fig.~\ref{fig:panel}, the horizontal range of each relaxation curve is chosen adaptively so that the relevant crossing and a short post-crossing interval remain visible; when $M_B=0$, the displayed range is restricted to $0\le\gamma_0t\le4$. All quoted crossing times and all values of $M_B$ are nevertheless computed over the full integration interval.

All two-dimensional parameter scans were evaluated on uniform $101\times101$ grids. For each parameter point, the master equation was integrated over the fixed interval $0\leq\gamma_0 t\leq20$, using the same temporal discretization for the evaluation of $M_B$. The representative relaxation curves shown in the figures were subsequently evaluated on a denser time grid in order to resolve the crossing region more clearly. Stationary states and time-dependent density matrices were obtained numerically with QuTiP, using the same Hamiltonian, collapse operators, and final Liouvillian defined above for both preparation protocols.

The master equation is solved by using the effective Hamiltonian in Eq.~\eqref{eq:Heff_source}, the thermal collapse operators of Eq.~\eqref{eq:thermalrates}, and the stationary states of the corresponding Liouvillians~\cite{Johansson2012,Johansson2013}. The reference numerical parameters are summarized in Table~\ref{tab:reference}. Throughout the parameter scans, the detuning ratio is kept fixed at $\Delta/\Omega=-1$. Accordingly, whenever $B$, $R$, $r$, or $E_0$ is varied, the drive frequency $\omega_d$ is adjusted point by point so that
\begin{equation}
\Delta=\omega(B,R,r)-\omega_d=-\Omega(E_0,B,R,r).
\end{equation}
Thus the detuning condition is preserved while the qubit transition frequency and the drive coupling change across parameter space. In the no-drive limit, $E_0\rightarrow0$ implies $\Omega\rightarrow0$ and therefore $\Delta\rightarrow0$, so that the prescription continuously reduces to the resonant undriven case.

\begin{table}[t]
\caption{Reference parameter set used in the numerical analysis unless
otherwise stated.}
\label{tab:reference}
\centering
\renewcommand{\arraystretch}{1.08}
\setlength{\tabcolsep}{7pt}
\begin{tabular}{@{}lc@{}}
\toprule
\multicolumn{2}{c}{\textit{Nanotorus geometry}}\\
\addlinespace[2pt]
Major radius, $R$       & $900~\text{\AA}$ \\
Minor radius, $r$       & $350~\text{\AA}$ \\
Magnetic field, $B$     & $0.45~\mathrm{T}$ \\
\addlinespace[4pt]

\multicolumn{2}{c}{\textit{Thermal protocol}}\\
\addlinespace[2pt]
Hot temperature, $T_h$     & $2~\mathrm{K}$ \\
Cold temperature, $T_c$    & $0.5~\mathrm{K}$ \\
Target temperature, $T_0$  & $0.05~\mathrm{K}$ \\
\addlinespace[4pt]

\multicolumn{2}{c}{\textit{Coherent control}}\\
\addlinespace[2pt]
Electric-field amplitude, $E_0$ & $10~\mathrm{V\,m^{-1}}$ \\
Drive phase, $\varphi$            & $0$ \\
Detuning ratio, $\Delta/\Omega$   & $-1$ \\
\addlinespace[4pt]

\multicolumn{2}{c}{\textit{Dissipation and numerics}}\\
\addlinespace[2pt]
Relaxation scale, $\gamma_0/2\pi$ & $1~\mathrm{MHz}$ \\
Pure dephasing, $\Gamma_\phi$      & $0$ \\
Integration window, $\gamma_0\tau$ & $20$ \\
\bottomrule
\end{tabular}
\end{table}

\section{Results}
\label{sec:results}

\subsection{Direct comparison at the reference point}
\label{subsec:reference}

Figure~\ref{fig:protocoldecay} compares the two preparation protocols at exactly the same final Hamiltonian, bath temperature, geometry, and magnetic field. For bare Gibbs preparation, the hotter state begins farther from the target, as required, but its distance decays sufficiently rapidly to cross the cold trajectory at
\begin{equation}
 \gamma_0t_{\rm cross}\simeq1.13.
\end{equation}
The resulting integrated parameter is
\begin{equation}
 M_B^{\rm G}\simeq0.661.
 \label{eq:MBgref}
\end{equation}
The associated $1/e$ distance-relaxation times are $\gamma_0\tau_h\simeq1.354$ and $\gamma_0\tau_c\simeq1.536$, consistent with the hot state relaxing faster despite starting farther from the target.

For the driven steady-state protocol, the behavior changes qualitatively. The two distances decay almost in parallel over the visible time interval and no early crossing is observed. The integrated value is only
\begin{equation}
 M_B^{\rm ss}\simeq2.47\times10^{-3},
 \label{eq:MBssref}
\end{equation}
with a crossing delayed until $\gamma_0t_{\rm cross}\simeq11.1$. The two $1/e$ relaxation times are practically identical, $\gamma_0\tau_h\simeq\gamma_0\tau_c\simeq1.333$. Thus the preparation protocol changes not merely the quantitative strength of the QMpE but the time scale on which the ordering reversal occurs.

\begin{figure*}[ht!]
    \centering
    \includegraphics[width=\textwidth]{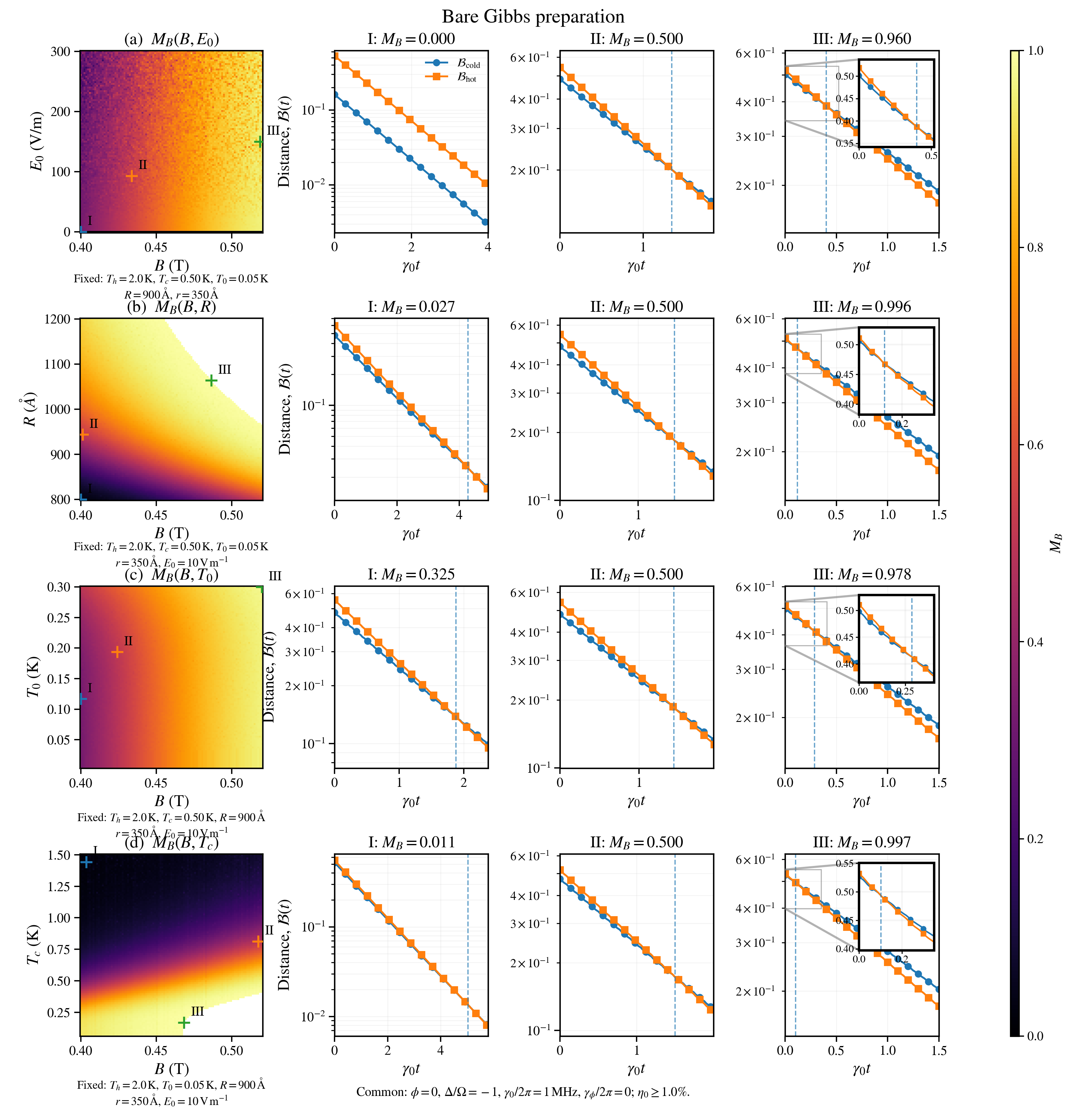}
    \caption{Summary of the bare-Gibbs preparation regime. The four rows show
    $M_B(B,E_0)$, $M_B(B,R)$, $M_B(B,T_0)$, and $M_B(B,T_c)$, respectively.
    In each row, the left panel displays the two-dimensional map, while the
    three panels to the right show the hot and cold distances for the
    representative points I, II, and III marked on the map. The integration
    defining $M_B$ is always performed over $0\le\gamma_0t\le20$, whereas the
    displayed time window is adapted to show the relevant crossing and a short
    post-crossing interval; when $M_B=0$, the plotted interval is
    $0\le\gamma_0t\le4$. Isolated numerical gaps in the maps are locally
    interpolated for visualization only and are not used in the quantitative
    analysis. The remaining parameters are fixed to the values indicated in
    each row, with common numerical settings listed at the bottom of the panel.}
    \label{fig:panel}
\end{figure*}

This result makes the role of preparation especially transparent. The final evolution operator $\exp(\Lio_0t)$ is identical in the two cases, so the difference must originate in the initial-state decomposition over the decay modes of $\Lio_0$. Bare Gibbs preparation followed by activation of the drive creates a state that is not stationary with respect to the driven generator, thereby generating a different population--coherence content than the driven stationary preparation.

\subsection{Electric-drive dependence}
\label{subsec:E0}

\begin{figure}[t]
\centering
\includegraphics[width=\columnwidth]{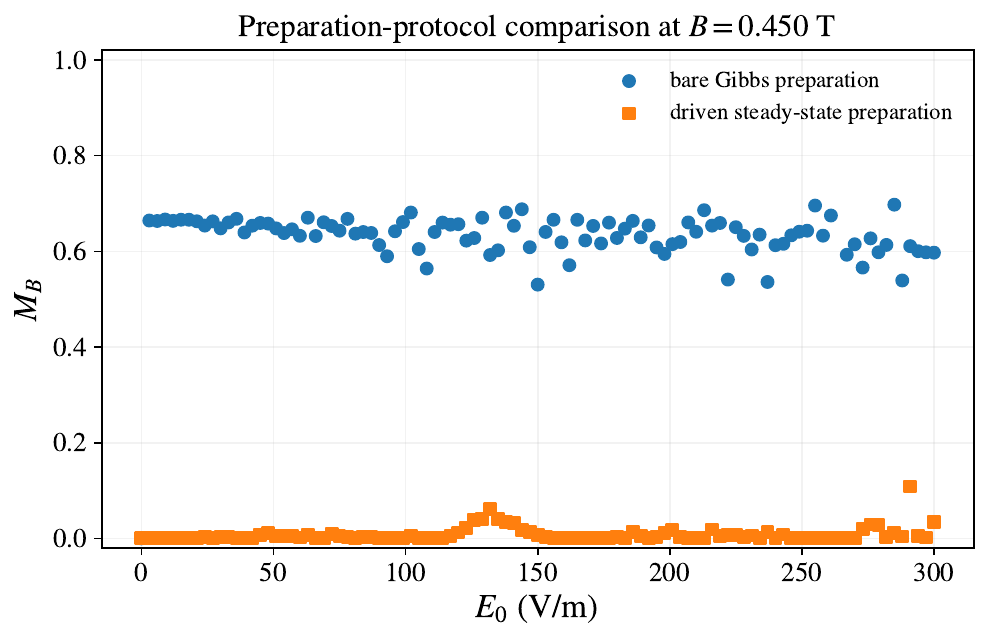}
\caption{$M_B$ as a function of electric-field amplitude for the two preparation protocols. The no-drive point is exactly zero in both cases. Bare Gibbs preparation rapidly develops a large QMpE once the drive is switched on, whereas driven steady-state preparation remains weak over most of the scan.}
\label{fig:MBprotocolsE0}
\end{figure}

The contrast persists when the electric-field amplitude is varied. Figure~\ref{fig:MBprotocolsE0} shows $M_B(E_0)$ for the two protocols at the reference values of all remaining parameters. Both curves correctly start at $M_B=0$ when $E_0=0$. However, already at the smallest nonzero fields sampled, the bare Gibbs protocol enters a broad plateau with $M_B\sim0.6$--$0.67$, and the effect remains large over essentially the whole interval explored. The driven steady-state protocol stays close to zero over most of the same range and becomes appreciable only at the largest electric fields.

The two-dimensional scans reinforce this separation. Rather than presenting
each bare-Gibbs map as an independent figure, Fig.~\ref{fig:panel} collects the
most informative parameter planes into a single summary panel. Each row contains
the corresponding $M_B$ map together with the distance dynamics at three
representative points, labelled I, II, and III, chosen to illustrate weak,
intermediate, and strong QMpE responses. This arrangement makes the relation
between the location in parameter space and the associated relaxation crossing
directly visible.

In the first row of Fig.~\ref{fig:panel}, corresponding to the
$B$--$E_0$ plane, the bare-Gibbs protocol displays a broad region of sizeable
QMpE once the coherent drive is switched on. The representative points evolve
from the no-drive limit, where $M_B=0$, through an intermediate regime with
$M_B\simeq0.48$, to a strong-response point with $M_B\simeq0.96$. As the effect
strengthens, the crossing moves toward earlier times, reaching
$\gamma_0t_{\rm cross}\simeq0.39$ at the strongest representative point.

The remaining rows show that this behavior is not peculiar to the electric
drive. In the $B$--$R$ plane, the QMpE strengthens continuously across a broad
portion of the explored geometry and reaches values extremely close to unity.
The $B$--$T_0$ plane exhibits a similarly wide high-$M_B$ region, while the
$B$--$T_c$ scan shows a pronounced separation between weak- and strong-response
regions as the preparation temperature is varied. In all three cases, the
distance panels provide the corresponding dynamical interpretation: increasing
$M_B$ is accompanied by an earlier reversal of the hot/cold ordering.

The steady-state preparation maps are not reproduced as an analogous summary
panel because they contain little extended QMpE structure on the common
$0\le M_B\le1$ scale. Their strongest values and associated crossing times are
instead summarized together with the bare-Gibbs results in
Table~\ref{tab:maxima}. The contrast is especially clear in the $B$--$E_0$
plane, where the strongest steady-state point reaches only
$M_B\simeq0.147$ with $\gamma_0t_{\rm cross}\simeq6.34$, compared with the
broad high-$M_B$ region obtained under bare-Gibbs preparation. It is important to 
highlight here that the white regions in the panel are associated with regions
outside the range of validity of the torus qubit encoding and/or controlling.

The two scans not included in Fig.~\ref{fig:panel}, namely the $B$--$T_h$ and
$B$--$r$ planes, follow the same qualitative hierarchy and are included in
Table~\ref{tab:maxima} for completeness. Under bare-Gibbs preparation,
$M_B$ approaches unity in both cases, whereas the corresponding steady-state
maxima remain at the percent level. Thus the panel in Fig.~\ref{fig:panel}
captures the representative structure of the bare-Gibbs regime without
requiring a sequence of largely redundant stand-alone maps.

\begin{table*}[t]
\caption{Largest sampled $M_B$ in each two-dimensional scan and the associated crossing time. These values correspond to the automatically selected point III in each map.}
\label{tab:maxima}
\begin{ruledtabular}
\begin{tabular}{lcccc}
Scan & $M_B^{\rm G}$ & $\gamma_0t_{\rm cross}^{\rm G}$ & $M_B^{\rm ss}$ & $\gamma_0t_{\rm cross}^{\rm ss}$\\
\hline
$(B,E_0)$ & $0.960$ & $0.390$ & $0.147$ & $6.34$\\
$(B,T_h)$ & $1.000$ & $1.0\times10^{-3}$ & $0.052$ & $8.25$\\
$(B,T_c)$ & $1.000$ & $0.015$ & $0.014$ & $9.33$\\
$(B,T_0)$ & $0.975$ & $0.302$ & $0.0033$ & $10.69$\\
$(B,R)$ & $1.000$ & $0.0044$ & $0.0032$ & $10.68$\\
$(B,r)$ & $1.000$ & $0.0014$ & $0.0035$ & $11.02$\\
\end{tabular}
\end{ruledtabular}
\end{table*}

\section{Discussion}
\label{sec:discussion}

\subsection{Why preparation matters: slow-sector analysis}

The preparation dependence can be understood directly in the Liouvillian eigenmode picture. Let
\begin{equation}
 \Lio_0|R_n\rangle\rangle=\lambda_n|R_n\rangle\rangle,
 \qquad
 \langle\langle L_n|\Lio_0=\lambda_n\langle\langle L_n|,
\end{equation}
with biorthogonal normalization. Any initial state evolves as
\begin{equation}
 |\rho(t)\rangle\rangle
 =|\rho_0\rangle\rangle
 +\sum_{n\neq0}\gamma_n\ee^{\lambda_nt}|R_n\rangle\rangle,
 \label{eq:modeexpansion}
\end{equation}
where
\begin{equation}
 \gamma_n=\langle\langle L_n|\rho(0)\rangle\rangle.
 \label{eq:gamma}
\end{equation}
The final eigenvalues $\lambda_n$ are identical for the two preparation protocols at the same physical parameter point because the final generator is identical. The preparation dependence can therefore enter only through the initial projections onto the decay modes.

For a unique slow mode, a natural diagnostic is
\begin{equation}
 R_s=\frac{|\gamma_{\rm slow}^{(h)}|}{|\gamma_{\rm slow}^{(c)}|},
 \label{eq:Rs}
\end{equation}
so that $R_s<1$ means that the initially hot state carries less weight than the cold state along the slowest active decay direction. In the driven qubit, however, the slow sector is generically two dimensional because the relevant Liouvillian modes form a decay-rate-degenerate pair. In that case an individual eigenvector inside the slow subspace is not unique, and Eq.~\eqref{eq:Rs} must be implemented in a basis-independent form.

Let $\mathsf{R}_s$ and $\mathsf{L}_s$ collect the right and left eigenvectors belonging to the slow active spectral sector. We define the corresponding spectral projector as
\begin{equation}
 \mathcal{P}_s=
 \mathsf{R}_s
 \left(\mathsf{L}_s^\dagger\mathsf{R}_s\right)^{-1}
 \mathsf{L}_s^\dagger,
 \label{eq:slowprojector}
\end{equation}
and the slow-sector weight of the hot or cold initial deviation by
\begin{equation}
 W_s^{(\nu)}=
 \left\|
 \mathcal{P}_s
 |\rho_\nu(0)-\rho_0\rangle\rangle
 \right\|_2,
 \qquad \nu\in\{h,c\}.
 \label{eq:slowweight}
\end{equation}
The numerical ratio used below is therefore
\begin{equation}
 R_s=\frac{W_s^{(h)}}{W_s^{(c)}}.
 \label{eq:Rsprojector}
\end{equation}
For a nondegenerate slow mode, with the right eigenvector normalized in Hilbert--Schmidt norm, Eq.~\eqref{eq:Rsprojector} reduces to Eq.~\eqref{eq:Rs}. We identify the \emph{slow active} sector as the slowest spectral sector having non-negligible projection on at least one of the two initial deviations. This excludes symmetry-protected modes that are spectrally slow but dynamically absent from both preparations.

The resulting maps show a systematic preparation-controlled reversal of the slow-sector weights. In the bare-Gibbs protocol, $\log_{10}R_s<0$ over essentially the whole driven QMpE region, whereas the driven steady-state protocol predominantly yields $\log_{10}R_s>0$. Thus the hot state is depleted in the slow active sector relative to the cold state for bare Gibbs preparation, but enriched in that sector for driven steady-state preparation. Since the final Liouvillian is the same at corresponding points, this contrast cannot be attributed to different decay spectra: it is a direct consequence of the initial-state preparation.

\begin{figure*}[t]
\centering
\includegraphics[width=0.98\textwidth]{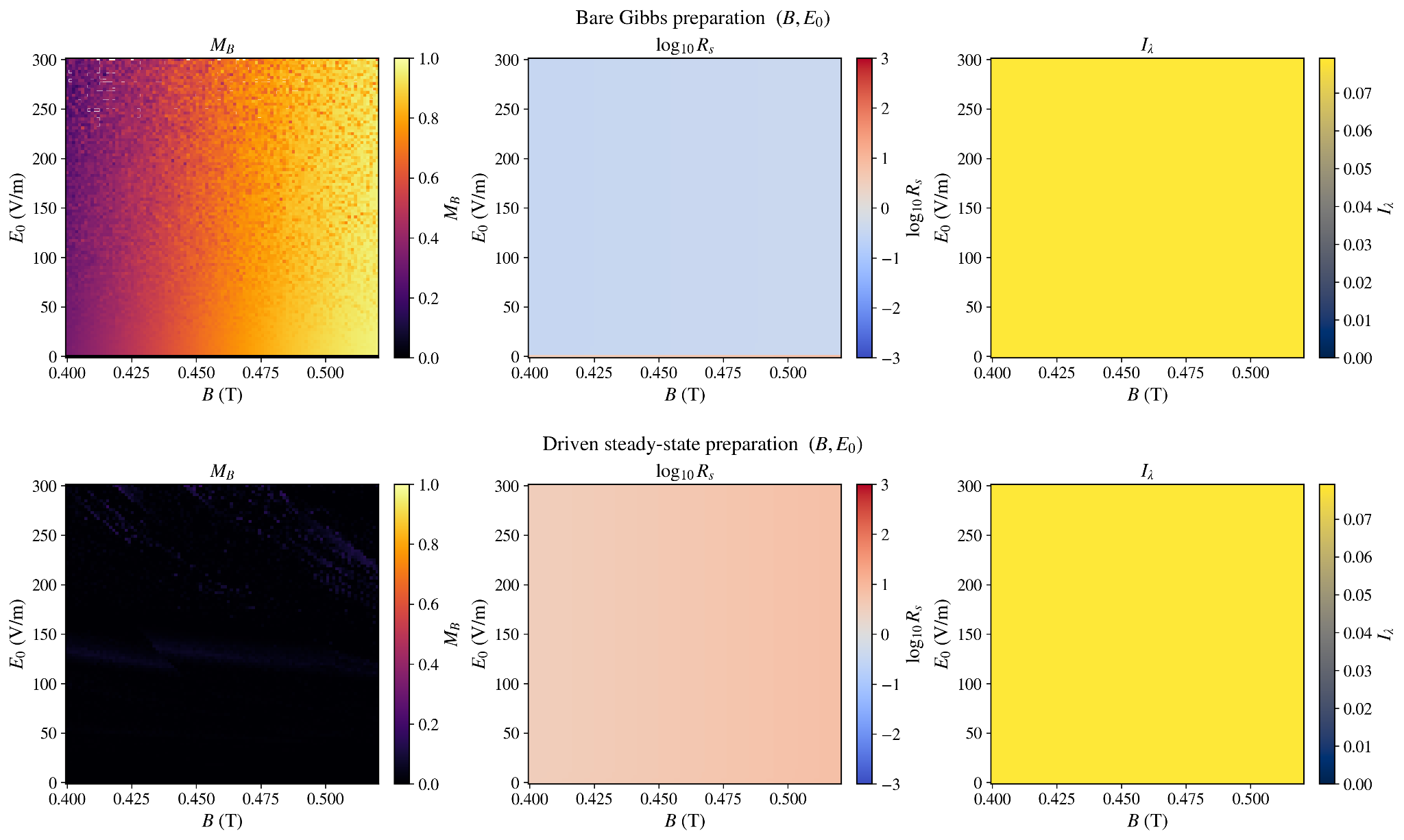}
\caption{Liouvillian-mode diagnostics in the $B$--$E_0$ plane. The upper block shows bare Gibbs preparation and the lower block driven steady-state preparation. In each block the panels display $M_B$, $\log_{10}R_s$, and the spectral-spread indicator $I_\lambda$. The same final Liouvillian is used at corresponding parameter points. Bare Gibbs preparation occupies the slow active sector with $R_s<1$ throughout the driven QMpE region, whereas driven steady-state preparation predominantly gives $R_s>1$ while $M_B$ remains close to zero. The comparatively weak variation of $I_\lambda$ shows that the protocol dependence is governed by initial-state modal weights rather than by a change of the final decay spectrum.}
\label{fig:RsBE0spectral}
\end{figure*}

Figure~\ref{fig:RsBE0spectral} illustrates this mechanism in the $B$--$E_0$ plane. For bare Gibbs preparation, switching on the coherent drive takes the system from the no-drive line with $M_B=0$ to a broad region with $R_s<1$ and sizeable $M_B$. The driven steady-state protocol exhibits the opposite modal ordering, with $R_s>1$ over almost the entire plane while $M_B$ remains close to zero. The spectral-spread indicator
\begin{equation}
 I_\lambda=
 \log_{10}\left(
 \frac{\Gamma_{\rm fast}}{\Gamma_{\rm slow}}
 \right),
 \qquad
 \Gamma_n=-\Re\lambda_n,
 \label{eq:Ilambda}
\end{equation}
varies only weakly compared with the large changes in $M_B$ and, at the same physical point, is necessarily identical for the two preparation protocols. The strong protocol dependence is therefore controlled primarily by modal populations rather than by a rearrangement of the Liouvillian decay-rate hierarchy.

\begin{figure*}[t]
\centering
\includegraphics[width=0.94\textwidth]{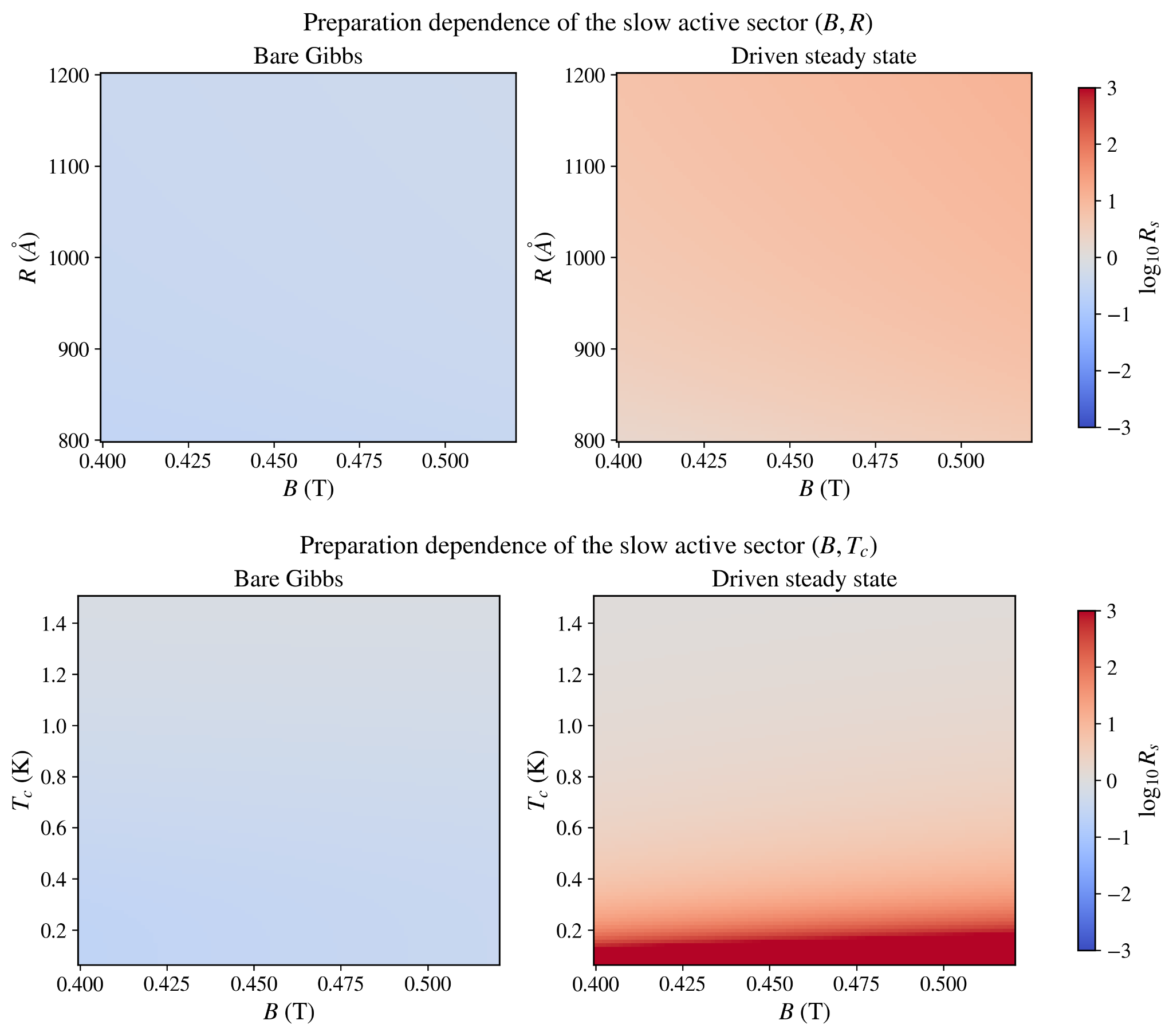}
\caption{Preparation dependence of $\log_{10}R_s$ for two representative scans. Top: $B$--$R$ plane. Bottom: $B$--$T_c$ plane. In both cases the bare-Gibbs protocol lies predominantly in the $R_s<1$ regime, while the driven steady-state protocol lies in the $R_s>1$ regime. The very large positive values at low $T_c$ in the steady-state protocol originate from strong suppression of the cold slow-sector weight and should therefore be interpreted as a small-denominator effect rather than an anomalous growth of the hot weight.}
\label{fig:Rsprotocolmaps}
\end{figure*}

The same separation persists away from the electric-drive scan. Figure~\ref{fig:Rsprotocolmaps} compares representative geometric and thermal planes. In the $B$--$R$ scan, the bare-Gibbs region remains uniformly on the $R_s<1$ side while the driven steady-state preparation remains on the $R_s>1$ side. In the $B$--$T_c$ plane, the contrast becomes particularly pronounced at low $T_c$, where the steady-state ratio can become very large. This large value should not be interpreted as an anomalously large absolute hot-state weight; rather, it signals a near suppression of the cold-state weight $W_s^{(c)}$ in the denominator of Eq.~\eqref{eq:Rsprojector}.

\begin{figure*}[t]
\centering
\includegraphics[width=0.98\textwidth]{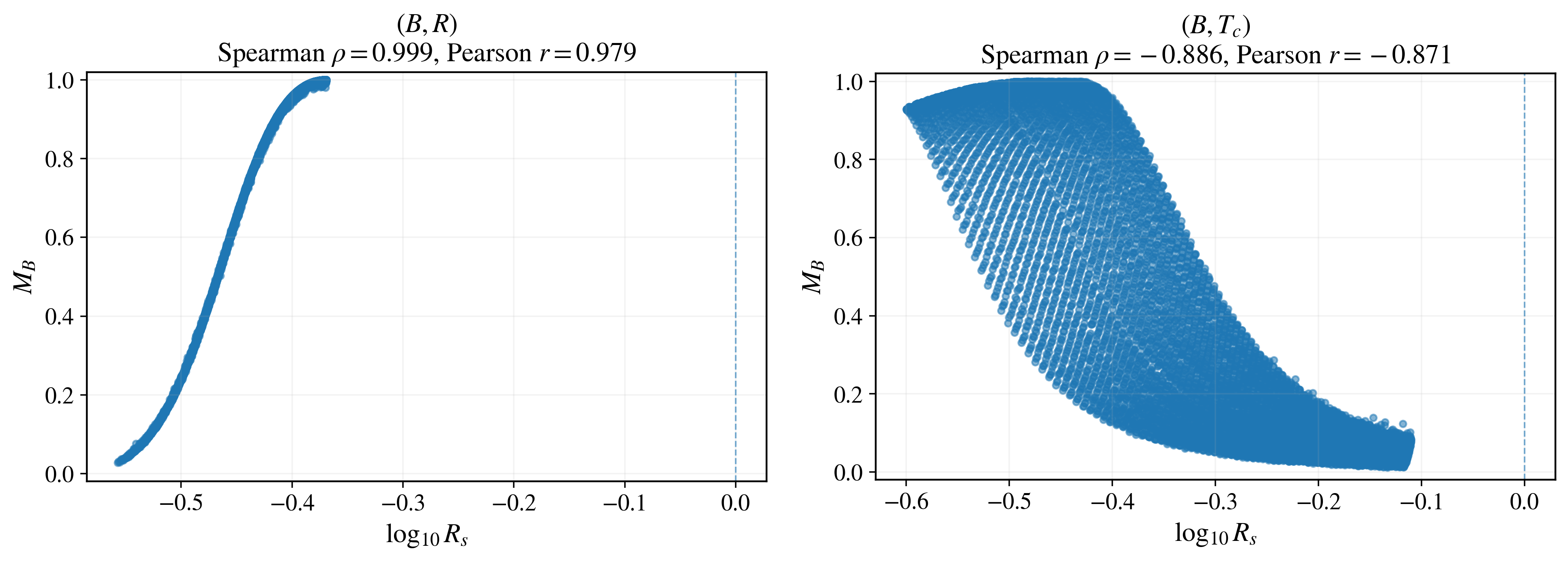}
\caption{Scan-dependent relation between the integrated Mpemba parameter and the slow-sector ratio for bare Gibbs preparation. Left: $B$--$R$ scan, showing an almost monotonic positive correlation. Right: $B$--$T_c$ scan, showing a strong negative correlation. The opposite trends demonstrate that $R_s<1$ identifies a modal condition favorable to the QMpE, but the numerical magnitude of $M_B$ is controlled by the complete multimode relaxation rather than by $R_s$ alone.}
\label{fig:Rscorrelations}
\end{figure*}

The relation between $R_s$ and the integrated Mpemba parameter is not universal in magnitude, even though the sign of $\log_{10}R_s$ cleanly distinguishes the two preparation regimes. Figure~\ref{fig:Rscorrelations} makes this point explicit for the bare-Gibbs protocol. In the $B$--$R$ scan, $M_B$ is almost monotonically increasing with $\log_{10}R_s$ (Spearman $\rho=0.999$, Pearson $r=0.979$), whereas in the $B$--$T_c$ scan the correlation is strongly negative (Spearman $\rho=-0.886$, Pearson $r=-0.871$). The $B$--$T_0$ and $B$--$T_h$ scans also show strong positive rank correlations, while the $B$--$E_0$ scan exhibits a distinct two-branch structure associated with the singular role of the no-drive line. Therefore $R_s<1$ should be interpreted as a favorable modal condition for QMpE, not as a universal order parameter for the value of $M_B$. The integrated quantity $M_B$ also depends on the weights and phases of faster modes, the initial distance ordering, and the time at which the crossing occurs.

\begin{figure}[t]
\centering
\includegraphics[width=\columnwidth]{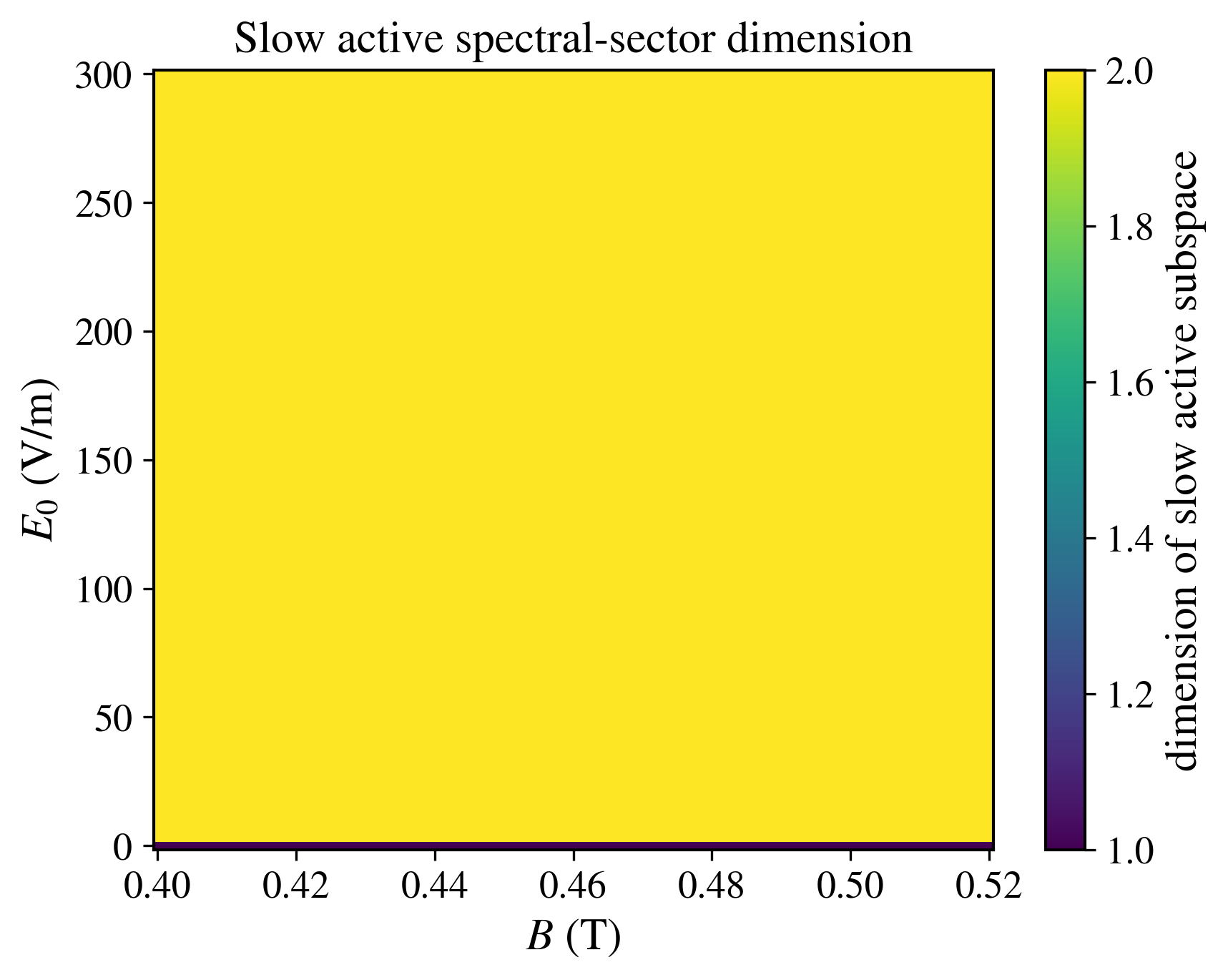}
\caption{Dimension of the slow active Liouvillian sector in the $B$--$E_0$ plane for bare Gibbs preparation. The active slow sector is one dimensional on the no-drive line $E_0=0$ and two dimensional throughout the driven region. The same final Liouvillian structure applies to the driven steady-state protocol at the corresponding physical parameters.}
\label{fig:slowdim}
\end{figure}

The dimensionality of the active slow sector provides an additional consistency check. For nonzero drive the slow active sector has dimension two throughout essentially the full scanned $B$--$E_0$ region, while it reduces to one on the $E_0=0$ line, as shown in Fig.~\ref{fig:slowdim}. This confirms that the projector formulation in Eq.~\eqref{eq:slowprojector} is the appropriate diagnostic for the driven qubit and also emphasizes the qualitative change introduced by coherent control: the no-drive limit is governed by a single active relaxation direction, whereas the driven problem explores a two-dimensional slow sector.

The steady-state scatter plots should be interpreted with additional caution. In most of those scans $M_B$ is numerically compressed near zero, so Pearson or Spearman coefficients are dominated by tiny residual variations and do not provide a useful measure of the underlying mechanism. The physically meaningful comparison is instead the protocol-level inversion of the slow-sector ordering: $R_s<1$ accompanies the broad bare-Gibbs QMpE regions, while $R_s>1$ accompanies the predominantly QMpE-suppressed driven steady-state regime. The calculation therefore converts the qualitative statement ``the initial states are different'' into a specific dynamical mechanism: the preparation protocol controls how the hot and cold states populate the slow decay sector of the same final Liouvillian.

\subsection{The strong bare-Gibbs response}

One of the most striking features of the scans is the breadth of the QMpE region under bare Gibbs preparation. This should not be interpreted as a contradiction with the relative rarity of Mpemba effects in generic relaxation. The preparation protocol contains an additional resource: at the quench, the system is not only moved to a colder bath, but also projected into the dynamics of a newly driven Hamiltonian. The coherent drive immediately couples population and coherence sectors, so the post-quench Liouvillian explores more than the single thermal population mode responsible for the no-drive result in Eq.~\eqref{eq:nocrossing}.

The electric-field scan supports this interpretation. Exactly at $E_0=0$, $M_B=0$, the slow active sector is one dimensional, and the favorable bare-Gibbs ordering $R_s<1$ is absent. At the smallest nonzero drive sampled, the active slow sector becomes two dimensional, the bare-Gibbs preparation moves to $R_s<1$, and the QMpE becomes large over a broad interval. Thus the model does not indicate an extended threshold region in $E_0$ at the current resolution; instead, the coherent quench rapidly changes both the active-mode structure and the initial-state projections onto it. Whether an experimentally realistic microscopic bath preserves this broad response is an important question for the next stage.

\subsection{Limitations and microscopic outlook}

The present results are obtained with phenomenological thermal rates. Therefore the maps should be interpreted as a controlled open-system model rather than a quantitative prediction of relaxation times in a fabricated nanotorus. A microscopic treatment should project the relevant graphene electron--phonon coupling onto the curvature-confined qubit states and determine
\begin{equation}
 \gamma(B,R,r)\propto
 J_{B,R,r}[\omega(B,R,r)]
 \left|\bra{0}\hat O_{\rm e-ph}\ket{1}\right|^2.
\end{equation}
The geometry can then affect not only the Hamiltonian but also the dissipative spectral density and transition matrix elements.

The effective two-level window nevertheless requires explicit control when the geometric or magnetic parameters are varied. In the present scans this consistency condition is imposed at the outset: the intervals of $B$, $R$, and $r$ are restricted to the qubit-admissible regime established in Ref.~\cite{Furtado:2022uvk}, where the $|l=1,m=0\rangle$ state is bound while additional $m=0$ bound states remain outside the computational sector. The two-level encoding is therefore preserved throughout the parameter domain considered here. Extensions beyond these intervals would require a new inspection of the full nanotorus spectrum.

Finally, the steady-state protocol does not eliminate the QMpE completely. It produces small but finite $M_B$ in some regions, especially at large electric fields. These late crossings may become more or less prominent once microscopic dephasing and electron--phonon rates are included. Their existence is useful because it shows that preparation acts as a continuous control of the relaxation hierarchy rather than as a binary ``on/off'' label.

\section{Conclusions}
\label{sec:conclusion}

We have studied quantum Mpemba relaxation in a graphene nanotorus qubit using the previously established magnetic confinement and electric-field control scheme. The no-drive limit reproduces the expected single-mode thermal relaxation and gives $M_B=0$. Once the coherent drive is present, the dynamics becomes strongly sensitive to how the hot and cold initial states are prepared.

Bare Gibbs preparation, in which the drive is activated at the final thermal quench, produces a robust QMpE over wide regions of the explored parameter space. At the reference point, the initially hotter state crosses the colder trajectory at $\gamma_0t\simeq1.13$ and yields $M_B\simeq0.661$. In several two-dimensional scans, $M_B$ approaches unity and the crossing moves to very early times. Driven steady-state preparation gives a dramatically different result under the same final Liouvillian: the reference value falls to $M_B\simeq2.5\times10^{-3}$, the crossing is delayed to $\gamma_0t\simeq11.1$, and most parameter maps remain close to zero.

The key conclusion is therefore not simply that the nanotorus can display a QMpE, but that the effect is highly \emph{preparation-protocol dependent}. The Liouvillian-mode decomposition makes this statement quantitative. Over the driven QMpE region, bare Gibbs preparation systematically places the hot state below the cold state in the weight carried by the slow active sector, $R_s<1$. Driven steady-state preparation predominantly reverses this ordering, $R_s>1$, while leaving the final spectrum unchanged at corresponding physical parameters. The weak variation of the spectral-spread indicator compared with the large changes in $M_B$ further shows that the dominant control mechanism is the preparation-dependent occupation of the decay modes rather than a restructuring of the Liouvillian eigenvalue hierarchy.

At the same time, the scan-dependent correlations between $M_B$ and $R_s$ show that the slow-sector ratio is not a universal order parameter for the magnitude of the effect. It provides a favorable modal condition, while the integrated $M_B$ also depends on the complete multimode content and crossing dynamics. The remaining decisive microscopic step is therefore not another phenomenological mode analysis but a device-specific dissipative treatment. Projecting electron--phonon coupling onto the curvature-confined qubit states will determine $\gamma(B,R,r)$ and establish whether the strong bare-Gibbs regions survive once the bath itself acquires the geometry dependence of the nanotorus.

\begin{acknowledgments}
JF would like to thank the Conselho Nacional de Desenvolvimento Científico e Tecnológico (CNPq) under grant 304485/2023-3, Alexandra Elbakyan and Sci-Hub, for removing all barriers in the way of science. 
\end{acknowledgments}

\section*{References}

\bibliography{mybib-URL.bib}

\end{document}